# Vortex and anti-vortex compositions of exact elegant Laguerre-Gaussian vector beams

**W. Nasalski**

Institute of Fundamental Technological Research,
Polish Academy of Sciences,
Adolfa Pawińskiego 5b, 02-106 Warsaw, Poland
e-mail: wnasal@ippt.pan.pl

**Abstract** Reformulation of conventional beam definitions into their bidirectional versions and use of Hertz potentials make the beam fields exact solutions to Maxwell's equations. Vortex and anti-vortex compositions of two such co-axial elegant Laguerre-Gaussian vector beams of TM and TE polarization and of equal or opposite vortex topological charges are given in a closed analytic form. The beam field compositions consist of paraxial and nonparaxial transverse parts of their strength dependent on a paraxial parameter. Longitudinal components are independent of this parameter. The new solutions obtained may appear useful in modelling and tailoring of arbitrary vector beams.

## 1  Introduction

Light beams can transport angular momentum (AM) along their propagation direction [1]. In general, AM of vector beams is composed of two parts – spin angular momentum (SAM) and orbital angular momentum (OAM). SAM is associated with beam circular polarization and has one from two values $\pm\hbar$ per photon. OAM is associated with phase profiles of beam fields. For beams of cylindrical symmetry and with helical phase fronts, like those of Laguerre-Gaussian (LG) form, OAM has also discrete values of $\pm l\hbar$, where $\pm l$ ($l \geq 0$) are topological charges of vortices placed at beam axes. For beam propagation in free space both parts of AM are independent of each other and separately conserved [2]. However, in inhomogeneous or anisotropic media, for spatially nonhomogeneous or anisotropic beam intensity and polarization, and besides their possible coupling with matter, SAM and OAM are in general interrelated and can be even mutually inter-converted. The phenomena of the SAM-OAM coupling or their spin-to-orbital conversion are not only interesting by themselves but may find many photonic applications [3]. For these reasons spatial structures of vector beams are recently under intense study [4-8]. Several approaches for producing and tailoring vector beams were proposed and discussed. In particular, complex structures of vector LG beams and their superpositions were analysed and interrelations between their polarization and field vortex structures were indicated. However, for beams of transverse diameters close to a field wavelength, paraxial description of beam fields is no longer adequate and exact representation of them governed by a complete set of Maxwell's equations should be implemented instead. Such exact representations will be presented in this paper.

Recently, a new exact representation of optical wave packets was proposed in [9]. That approach, which covers also the field synthesis of exact beam solutions discussed in this report, consists of three subsequent steps. In the first step, time and a spatial coordinate along the beam propagation direction are treated on equal footing. That results in the bidirectional representation of beam fields, where beam envelopes depend not only on spatial coordinates but also on time. This technique, known in electomagnetics and optics for a long time [10,11], has been extensively used in construction, for example, localized wave packets [12-14]. Next, a scalar solution to the propagation problem is stipulated in the modified form of the elegant (complex argument) Laguerre-Gaussian (eLG) beams. Their conventional form is known in their paraxial version where they constitute a complete and bi-orthogonal base of square integrated functions [15-16]. Basic definitions and characteristic features of such beams were recently presented in [17] in the context of their interactions with dielectric interfaces. However, because of the different definition of the beam envelope introduced in the first step and due to field propagation factors, the scalar eLG beam fields are now exact, not paraxial, as they obey exactly the wave equation. In the final step, a full vector representation of the beam field is given in terms of Hertz vector potentials [10]. Orthogonal, transverse with respect to the beam axis, magnetic (TM) and electric (TE) exact solutions to Maxwell's equations can be then readily obtained.

One new element in this construction is the bidirectional modification of the monochromatic eLG beam fields and systematic use of them in the construction of higher-order vector solutions. The eLG beams are well behaved physical entities – they carry finite energy, linear momentum and angular momentum per unit length along their propagation direction. Their symmetries are particularly suitable in evaluation cross-polarization effects at planar interfaces and multilayers [9,17]. They can always be given in a closed form in both, on a parallel, configuration and spectral domains. Thus, no approximations or expansions are necessary in the analysis this type. The field representation is given explicitly in an analytic form with a beam spatial shape determined by indices of the eLG functions and a diameter of a beam cross section.

Another new element in this analysis is the explicit separation of exact beam fields into two ingredients recognized as nonparaxial and paraxial parts of exact solutions. They are also exact scalar solutions to the wave equation by themselves and are distinguished by a factors expressed by a paraxial parameter and its inverse, respectively. This field separation can be accomplished by appropriate scaling of spatial and spectral coordinates in vector representations of the beam fields. Hence, in all expressions, except these in Eqs (12)-(13), in which the propagation direction is not specified, the spatial and the space-time coordinates are used in normalized, dimensionless forms.

The transverse coordinates $x$ and $y$ are normalized to (divided by) a transverse scale $w_w$ of the meaning of a beam radius at its waist. The longitudinal (along the propagation direction) coordinates $z$ and $\tau = ct$, as well as the phase front curvature radius $R$, are normalized to (divided by) $z_D = k w_w^2$ of the meaning of a beam diffraction or Rayleigh length; $k$, $c$ and $t$ are a wave number, phase velocity and time, respectively. The ratio $f = 2^{-1/2} w_w z_D^{-1}$ of the both scales determines a beam paraxiality level. The wave number $k$ is also normalized to $z_D$. That leads also to the identity $k^2 = z_D^2 w_w^{-2}$. Similarly, in the spectral domain, the transverse $k_x$, $k_y$ coordinates are normalized to (multiplied by) $w_w$. The same normalization concerns also the alternative, longitudinal (real) and transverse (complex) coordinates in the configuration and spectral domains, respectively:

$$z_\pm = z \pm \tau, \qquad \varsigma_\pm = \tfrac{1}{\sqrt{2}}(x \pm iy), \qquad \kappa_\pm = \tfrac{1}{\sqrt{2}}(k_x \pm i k_y). \tag{1}$$

Equivalent expressions $\varsigma_\pm = \varsigma_\perp e^{\pm i\phi}$ and $\kappa_\pm = \kappa_\perp e^{\pm i\varphi}$ for the transverse coordinates relate them to the vortices of the topological charges $\pm 1$ and to the angular coordinates $\phi$ and $\varphi$ defined by $\tan\phi = yx^{-1}$ and $\tan\varphi = k_y k_x^{-1}$ in the transverse $x-y$ and $k_x - k_y$ planes, respectively. All beams considered here are of cylindrical symmetry with the same beam axis along the z-axis and with their waists of the same diameter $2w_w$ placed in the common transverse plane $z=0$. Additional normalization of the electromagnetic field vectors will be also introduced in section 3.

In this paper, construction of exact eLG solutions is presented for beams propagating in an unbounded, homogeneous, isotropic and transparent medium. Scalar and vector versions of eLG beams of arbitrary order will be analysed in sections 2 and 3, respectively. Vortex and anti-vortex superpositions of two overlapping and co-propagating beams will be presented in section 4 and 5, respectively. The analysis will be concluded in section 6.

## 2  Exact scalar eLG beams

Let us start from the scalar monochromatic field $g'$ obeying the wave equation in free space:

$$[2(w_w/z_D)^2 \partial_{z_+}\partial_{z_-} + \partial_{\varsigma_+}\partial_{\varsigma_-}]g'(\varsigma_+,\varsigma_-,z_+,z_-) = 0. \quad (2)$$

The function $g'$ depends on both longitudinal coordinates $z_\pm$ and on the specified wave number $k$. If $g'$ is factorised into its envelope $g$ dependent on $z_+$ and the propagation factor dependent on $z_-$; $g' = g\exp(ikz_-)$, then the wave equation reduces exactly to the paraxial equation with its fundamental Gaussian solution $g$ [17]:

$$(2i\partial_{z_+} + \partial_{\varsigma_+}\partial_{\varsigma_-})g(\varsigma_+,\varsigma_-,z_+) = 0, \quad (3)$$

$$g(\varsigma_+,\varsigma_-,z_+) = v^{-2}(z_+)\exp[-\varsigma_+\varsigma_- v^{-2}(z_+)]. \quad (4)$$

The beam complex radius $v$ specifies the real beam parameters: the beam width diameter $2w$ and the radius $R$ of phase front curvature; $v^2 = 1 + i2^{-1}z_+ = (w^{-2} - iR^{-1})^{-1}$. On the other hand, the paraxial equation (3) describes also the conventional Gaussian envelope $\breve{g}$ after the replacement of $z_+$ by $2z$, as its complex width squared $\breve{v}^2(z) = 1 + iz$ depends only on $z$.

Higher-order exact solutions $G_{p,\pm l}$ to (3) are obtained by the appropriate differentiation of the Gaussian beam $g \equiv G_{0,0}$ [17]. That separates the vortex factor; $G_{p,\pm l} = Q_{p,l}\exp(\pm il\phi)$, where

$$G_{p,\pm l}(\varsigma_+,\varsigma_-,z_+) = \partial_{\varsigma_\pm}^p \partial_{\varsigma_\mp}^{p+l} g(\varsigma_+,\varsigma_-,z_+), \quad (5)$$

$$Q_{p,l}(\varsigma_+,\varsigma_-,z_+) = (-1)^{p+l} p! v^{-(2p+l)} u^l L_p^l(u^2) g(\varsigma_+,\varsigma_-,z_+), \quad (6)$$

$u^2 = \varsigma_+\varsigma_- v^{-2}$, $l = |l|$ and $Q_{0,0} = g$. The function $Q_{p,l}$ and the associated Laguerre polynomials $L_p^l$ are specified by magnitudes of the radial azimuthal indices $p$ and $l$, respectively. They are independent of $\phi$ and the sign of $\pm l$. Note that the conventional paraxial eLG beams $\breve{G}_{p,\pm l}$ are also defined by (5)-(6) with the single replacement $z_+$ by $2z$.

The definitions (5)-(6) can be also restated in the spectral domain through the 2D Fourier transform in the transverse planes $\varsigma_+ - \varsigma_-$ and $\kappa_+ - \kappa_-$ [17]:

$$G_{p,\pm l}(\varsigma_+,\varsigma_-,z_+) = \tfrac{i}{2\pi}\int d\kappa_+ d\kappa_- \tilde{G}_{p,\pm l}(\kappa_+,\kappa_-,z_+)e^{i(\kappa_+\varsigma_- + \kappa_-\varsigma_+)}. \quad (7)$$

$$\tilde{G}_{p,\pm l}(\kappa_+,\kappa_-,z_+) = i^{2p+l}\kappa_\mp^p \kappa_\pm^{p+l}\tilde{g}(\kappa_+,\kappa_-,z_+), \quad (8)$$

$$\tilde{Q}_{p,l}(\kappa_+,\kappa_-,z_+) = (-\kappa_+\kappa_-)^{p+l/2}\tilde{g}(\kappa_+,\kappa_-,z_+), \quad (9)$$

where $\tilde{G}_{p,\pm l} = \tilde{Q}_{p,l}\exp(\pm il\varphi)$ and $\tilde{g} = \exp(-\kappa_+\kappa_- v^2)$. The definition (8) allows also for the alternative representations of the eLG functions:

$$\tilde{G}_{p,\pm l} = \tilde{G}_{p\pm 1/2,\pm(l-1)}e^{\pm i\varphi} = \tilde{G}_{p\pm 1,\pm(l-2)}e^{\pm 2i\varphi}, \quad (10)$$

$$\tilde{G}_{p,\pm l} = \tilde{G}_{p\mp 1/2,\pm(l+1)}e^{\mp i\varphi} = \tilde{G}_{p\mp 1,\pm(l+2)}e^{\mp 2i\varphi}, \quad (11)$$

in the spectral domain [17] and in parallel for $G_{p,\pm l}$ in the configuration domain [9], where $\tilde{G}_{p,\pm l}$ and $G_{p,\pm l}$ with their indices $p$, $\pm l$ and the azimuthal angles $\varphi$ and $\phi$, respectively, are interrelated by the Fourier transform (7). In derivations of vector solutions polarization symmetries came into play as well together with these interrelations (cf. [9] and Eqs (16)-(17) in section 3).

In both, scalar and vector, beam solutions, their radii $v$, $w$ and $R$ depend on $z_+$ and in turn $z_+$ depends on both, longitudinal and time, spatial coordinates $z$ and $\tau = ct$. Therefore, the eLG beams $G_{p,\pm l}$ differ in general from the conventional paraxial eLG beams $\breve{G}_{p,\pm l}$. However, at the initial time $t = 0$, $z_+ = z$ and the beams copy their conventional counterparts, although with its coordinate $z$ shortened twice. Moreover, at the phase front plane and for any time moment, $z = ct$ or $z_- = 0$ and $z_+ = 2z$. In this case the beam field distribution remains exactly in the conventional form with the following changes: its waist position (at $z_+ = 0$) is shifted along z-axis by $-ct$, together with its on-axis phase (at $kz_- = 0$) shifted along $z$ by $-2kz$. Similar relations for the beam on-axis phase shift and waist position can be found for arbitrary values of $t$ and $z$. Thus the exact eLG beams are physical entities to the same extent as the conventional paraxial eLG beams are.

## 3  Exact vector eLG beams

Exact solutions to Maxwell's equations can be build from the Hertz vector potentials $\boldsymbol{M}'$ and $\boldsymbol{N}'$ [10]. For symmetry reasons both of them are taken directed along the beam axis, that is, they possess only one nonzero component $M'_z$ and $N'_z$, respectively. The total field can be then decomposed of the two collinear and orthogonal TM and TE solutions:

$$\boldsymbol{E}' = \boldsymbol{E}'^{(tm)} + \boldsymbol{E}'^{(te)} = \nabla \times \nabla \times \boldsymbol{M}' - \partial_\tau \nabla \times \boldsymbol{N}', \qquad (12)$$

$$\boldsymbol{H}' = \boldsymbol{H}'^{(te)} + \boldsymbol{H}'^{(tm)} = \nabla \times \nabla \times \boldsymbol{N}' + \partial_\tau \nabla \times \boldsymbol{M}', \qquad (13)$$

where, exceptionally, the coordinates $x$, $y$, $z$ and $t$ are not normalized in these equations. In (12)-(13) the electric $\boldsymbol{E}'$ and magnetic $\boldsymbol{H}'$ field vectors are normalized (multiplied) by square roots of an intrinsic admittance $Y$ and an impedance $Z$ of the medium, respectively. In the cylindrical circular polarization frame $(\hat{\boldsymbol{e}}_R, \hat{\boldsymbol{e}}_L, \hat{\boldsymbol{e}}_z)$ and with the potentials $M'_z = M'w_w^2$ and $N'_z = N'w_w^2$, Eqs (12)-(13) yield [9]:

$$\boldsymbol{E}'^{(tm)} = -2\hat{\boldsymbol{e}}_z \partial_{\varsigma_+} \partial_{\varsigma_-} M' + (w_w/z_D)(\hat{\boldsymbol{e}}_R \partial_{\varsigma_+} + \hat{\boldsymbol{e}}_L \partial_{\varsigma_-})\partial_z M', \qquad (14)$$

$$\boldsymbol{E}'^{(te)} = -i(w_w/z_D)(\hat{\boldsymbol{e}}_R \partial_{\varsigma_+} - \hat{\boldsymbol{e}}_L \partial_{\varsigma_-})\partial_\tau N', \qquad (15)$$

where $\partial_z = \partial_{z_+} + \partial_{z_-}$ and $\partial_\tau = \partial_{z_+} - \partial_{z_-}$. The TM and TE solutions are in general independent. If, however, $N'_z = \pm M'_z$, then by duality $\boldsymbol{H}'^{(tm)} = \pm \boldsymbol{E}'^{(te)}$ and $\boldsymbol{H}'^{(te)} = \mp \boldsymbol{E}'^{(tm)}$. Let the Hertz scalars $M'$ and $N'$ and the field vectors $\boldsymbol{E}'^{(tm)}$ and $\boldsymbol{E}'^{(te)}$ represent by the envelope factors $M$, $N$, $\boldsymbol{E}^{(tm)}$ and $\boldsymbol{E}^{(te)}$ dependent on $z_+$ and the common propagation factor $\exp(ikz_-)$. Consider first the case of equal Hertz potentials expressed by the eLG function $N = M = w^2 G_{p,\pm l}$ and covert (14)-(15) from the frame $(\hat{\boldsymbol{e}}_R, \hat{\boldsymbol{e}}_L, \hat{\boldsymbol{e}}_z)$ to the cylindrical polar (radial/azimuthal) polarization frame $(\hat{\boldsymbol{e}}_\rho, \hat{\boldsymbol{e}}_\phi, \hat{\boldsymbol{e}}_z)$:

$$\hat{\boldsymbol{e}}_\rho = 2^{-1/2}(\hat{\boldsymbol{e}}_R e^{-i\phi} + \hat{\boldsymbol{e}}_L e^{+i\phi}), \qquad (16)$$

$$i\hat{\boldsymbol{e}}_\phi = 2^{-1/2}(\hat{\boldsymbol{e}}_R e^{-i\phi} - \hat{\boldsymbol{e}}_L e^{+i\phi}). \qquad (17)$$

That yields the beam field decomposition into the TM and TE orthogonal components each consisted of the two independent solutions differentiated by their polarization and the sign of $\pm l$:

$$\boldsymbol{E}_{p,\pm l}^{(tm)} = -2\hat{\boldsymbol{e}}_z Q_{p+1,l} e^{\pm il\phi}$$
$$+i\hat{\boldsymbol{e}}_\rho (f^{-1}Q_{p,l+1}+f^{+1}Q_{p+1,l+1})e^{\pm il\phi}, \tag{18}$$

$$\boldsymbol{E}_{p,\pm l}^{(te)} = i\hat{\boldsymbol{e}}_\phi (f^{-1}Q_{p,l+1}-f^{+1}Q_{p+1,l+1})e^{\pm il\phi}, \tag{19}$$

In derivation of (18)-(19) the definitions of the paraxial equation (3) and the eLG beams (4)-(9) were used [9]. The magnetic field can be then obtained from the duality principle. Note that Eqs (18)-(19) are, on the grounds of the identities (10)-(11), equivalent to Eqs (9)-(10) in Ref. [9].

The representation (18)-(19) shows quite regular field structure - for the separate paraxial and nonparaxial parts the field spatial distribution appears of the same form for both transverse polarization components. The parameter $f$ indicates what part of the solution prevails for paraxial ($f \ll 1$) or nonparaxial ($f \gg 1$) values of the ratio of the beam waist radius to the field wavelength. The paraxial, longitudinal and nonparaxial contributions to the total field possess the same vortex factor $\exp(\pm il\varphi)$ specified by the azimuthal index $l$. Note that the paraxial and nonparaxial field ingredients in (18)-(19) satisfy exactly the paraxial equation (3) meanwhile the field polarization components in (14)-(15) satisfy exactly the wave equation (2).

## 4  Vortex composition of two co-axial vector eLG beams

For vortex compositions the scalar Hertz potentials in (14)-(15) are equal; $N=M$. Moreover, the definitions (18)-(19) of the radial and azimuthal polarization vectors suggest the alternative field representation in the cylindrical circular polarization hybrid frame ($\hat{\boldsymbol{e}}_R \exp(-i\phi), \hat{\boldsymbol{e}}_L \exp(+i\phi), \hat{\boldsymbol{e}}_z$). The transverse components of this frame have zero AM as their SAM are cancelled by their opposite OAM. The coherent superpositions with relative phase $\mp \pi/2$: $\boldsymbol{E}_{p,\pm l}^{(a)}=2^{-1/2}(\boldsymbol{E}_{p,\pm l}^{(tm)}-i\boldsymbol{E}_{p,\pm l}^{(te)})$ and $\boldsymbol{E}_{p,\pm l}^{(b)}=2^{-1/2}(\boldsymbol{E}_{p,\pm l}^{(tm)}+i\boldsymbol{E}_{p,\pm l}^{(te)})$, of the TM and TE vector eLG beams (18)-(19) of the same $p$ and $\pm l$ indices results in the new vector structure of vortex beams:

$$\boldsymbol{E}_{p,\pm l}^{(a)} = \boldsymbol{E}_{\perp;p,\pm l}^{(a)} - 2^{1/2}\hat{\boldsymbol{e}}_z Q_{p+1,l} e^{\pm il\phi}, \tag{20}$$

$$\boldsymbol{E}_{p,\pm l}^{(b)} = \boldsymbol{E}_{\perp;p,\pm l}^{(b)} - 2^{1/2}\hat{\boldsymbol{e}}_z Q_{p+1,l} e^{\pm il\phi}, \tag{21}$$

$$\boldsymbol{E}_{\perp;p,\pm l}^{(a)} = i(\hat{\boldsymbol{e}}_L e^{+i\phi} f^{-1} Q_{p,l+1} + \hat{\boldsymbol{e}}_R e^{-i\phi} f^{+1} Q_{p+1,l+1}) e^{\pm il\phi}, \tag{22}$$

$$\boldsymbol{E}_{\perp;p,\pm l}^{(b)} = i(\hat{\boldsymbol{e}}_R e^{-i\phi} f^{-1} Q_{p,l+1} + \hat{\boldsymbol{e}}_L e^{+i\phi} f^{+1} Q_{p+1,l+1}) e^{\pm il\phi}, \tag{23}$$

with the transverse field components labelled by the subscript $\perp$ are described by four independent formulae (differentiated by the sign of $\pm l$ in the cases (a) and (b)). The field structure is now even more regular than that one in the cylindrical polar polarization frame. For the separate − paraxial, longitudinal and nonparaxial - field contributions the exact field spatial structures (a) and (b) appear of the same spatial form. The transverse beam components (22)-(23) show symmetry with respect to the beam polarization; $\boldsymbol{E}_{p,\pm l}^{(a)}$ can be obtained from $\boldsymbol{E}_{p,\pm l}^{(b)}$, and vice versa, by the simple replacement $\hat{\boldsymbol{e}}_R \exp(-i\phi) \leftrightarrow \hat{\boldsymbol{e}}_L \exp(+i\phi)$. The longitudinal components are of the same form in any one from four solutions given in (20)-(23). The amplitude ratio between the nonparaxial part and the paraxial part of the field are always related by the ratio $f^{+2} Q_{p+1,l+1}/Q_{p,l+1}$ and OAM values of the all polarization components amount $\pm l\hbar$ per photon. In spite of the use of the elegant rather than the standard LG beams, the paraxial parts of the solution (22)-(23) correspond to the compositions of the paraxial solutions (I)-(IV) presented in [8] standard LG beams.

## 5  Anti-vortex composition of two co-axial vector eLG beams

In the case of anti-vortex compositions the scalar Hertz potentials in (14)-(15) take on the form $M = G_{p,\pm l}$ and $N = G_{p,\mp l}$, namely, they are of opposite azimuthal indices $\pm l$ and $\mp l$, respectively. In this case the TM solutions in (18) are the same as before but the TE solutions (19) now possess the different (opposite) vortex factor $\exp(\mp il\phi)$. The compositions are then obtained by the $\mp \pi/2$ phase shifted superpositions: $\boldsymbol{E}^{(c)}_{p,\pm l} = 2^{-1/2}(\boldsymbol{E}^{(tm)}_{p,\pm l} - i\boldsymbol{E}^{(te)}_{p,\mp l})$ and $\boldsymbol{E}^{(d)}_{p,\pm l} = 2^{-1/2}(\boldsymbol{E}^{(tm)}_{p,\pm l} + i\boldsymbol{E}^{(te)}_{p,\mp l})$, with the transverse field components represented by four independent solutions:

$$\boldsymbol{E}^{(c)}_{p,\pm l} = \boldsymbol{E}^{(c)}_{\perp;p,\pm l} - 2^{1/2}\hat{\boldsymbol{e}}_z Q_{p+1,l} e^{\pm il\phi}, \qquad (24)$$

$$\boldsymbol{E}^{(d)}_{p,\pm l} = \boldsymbol{E}^{(d)}_{\perp;p,\pm l} - 2^{1/2}\hat{\boldsymbol{e}}_z Q_{p+1,l} e^{\mp il\phi}, \qquad (25)$$

$$\begin{aligned}\boldsymbol{E}^{(c)}_{\perp;p,\pm l} = & \\ &+\hat{\boldsymbol{e}}_L e^{+i\phi}[if^{-1}\cos(l\phi)Q_{p,l+1} \mp f^{+1}\sin(l\phi)Q_{p+1,l+1}], \\ &\mp \hat{\boldsymbol{e}}_R e^{-i\phi}[f^{-1}\sin(l\phi)Q_{p,l+1} \mp if^{+1}\cos(l\phi)Q_{p+1,l+1}]\end{aligned} \qquad (26)$$

$$\begin{aligned}\boldsymbol{E}^{(d)}_{\perp;p,\pm l} = & \\ &+\hat{\boldsymbol{e}}_R e^{-i\phi}[if^{-1}\cos(l\phi)Q_{p,l+1} \mp f^{+1}\sin(l\phi)Q_{p+1,l+1}]. \\ &\mp \hat{\boldsymbol{e}}_L e^{+i\phi}[f^{-1}\sin(l\phi)Q_{p,l+1} \mp if^{+1}\cos(l\phi)Q_{p+1,l+1}]\end{aligned} \qquad (27)$$

The symmetries $\hat{\boldsymbol{e}}_R \exp(-i\phi) \leftrightarrow \hat{\boldsymbol{e}}_L \exp(+i\phi)$ between $\boldsymbol{E}^{(c)}_{\perp;p,\pm l}$ and $\boldsymbol{E}^{(d)}_{\perp;p,\pm l}$ are similar to those for the vortex composition but in (26)-(27) the trigonometric functions replace the corresponding exponent functions in (22)-(23) separately for the paraxial and nonparaxial parts of the solution. The intensity patterns of these anti-vortex compositions comprise $2l$ petals, placed in opposed angular positions for paraxial and nonparaxial parts. The amplitude ratios between the nonparaxial part and the paraxial part are now equal $\mp i f^{+2}\cot(l\phi)Q_{p+1,l+1}/Q_{p,l+1}$ and $\pm i f^{+2}\tan(l\phi)Q_{p+1,l+1}/Q_{p,l+1}$ for the cylindrical right-handed and left-handed circular polarization components of $\boldsymbol{E}^{(c)}_{\perp;p,\pm l}$, respectively, and vice versa for $\boldsymbol{E}^{(d)}_{\perp;p,\pm l}$. For the paraxial part of the solution (26)-(27), the correspondence can also be found with results presented in [6] for the standard LG beams.

## 6  Conclusions

In general, beam field compositions can be obtained by arbitrary superposition of TM and TE solutions $\alpha\boldsymbol{E}^{(tm)}_{p,\pm l} + \beta\boldsymbol{E}^{(te)}_{p,l'}$ with complex numbers $\alpha$ and $\beta$, where $l'=\pm l$ and $l'=\mp l$ for the vortex and anti-vortex compositions, respectively. For the solutions (20)-(27) $\beta=\pm i\alpha$. Other beam compositions, like these of $\beta=\pm \alpha$, i.e. without the phase shift $\pi/2$, can be derived per analogy. There are also other choices in construction vector eLG beams by using, for example, transverse instead of longitudinal vector potentials or collinear eLG beams of different amplitudes and/or with different vortex charges. However, the vortex and anti-vortex co-axial compositions presented here account fully for the cylindrical symmetry of the beam fields as they are built from the longitudinal and equal in magnitude components of Hertz potentials. Characteristics of these compositions seem useful for interpretation of more complex field structures of beams.

The solutions obtained are composed of two transverse - paraxial and nonparaxial - parts, which are always associated with the paraxial parameter and its inverse, respectively. On the contrary, the longitudinal field components do not depend on this parameter. Each of these parts satisfies the wave equation separately and their radial or azimuthal indices specify differently each part of the beam field. Although the circular polarization frame ($\hat{e}_R, \hat{e}_L$) appears to be convenient in the field derivations and the TM and TE modes are separated in the polar frame ($\hat{e}_\rho, \hat{e}_\phi$), the hybrid frame ($\hat{e}_R \exp(-i\phi), \hat{e}_L \exp(+i\phi)$) seems more suitable for description of the beam compositions.

It should be finally indicate that the beams considered here are monochromatic of forward propagation type. They are obtained for one specified value of the wave number $k$. However, the beam fields depend on $z_-$ only through their propagation factors meanwhile the field envelopes depend on $z_+$. Moreover, after the replacement $z_\pm \to z_\mp$ in all above expressions, the results presented here remain also valid in the backward propagation case. The field compositions of counter-propagating beams are then readily available by analogy to the case of co-propagating beams. In general, series consisted of the solutions of both these types, with different radial and azimuthal indices and polarization, yield beam fields determined by arbitrary initial conditions specified at any initial plane transverse to the beam propagation direction.

**References**

1. L. Allen, S. M. Barnett, and M. J. Padgett, *Orbital Angular Momentum* (Institute of Optics Publishing, 2003)
2. S. M. Barnett, J. Opt. B: Quantum Semiclass. Opt. **4**, S7 (2002)
3. L. Marrucci, E. Karimi, S. Slussarenko, B. Piccirillo, E. Santamato, E. Nagali and F. Sciarrino, J. Opt. **13**, 064001 (2011)
4. M. A. A. Neil, T. Wilson and T. Juškaitis, Journal of Microcopy **197**, 219 (2000)
5. C. Maurer, A. Jesacher, S. Fürhapter, S. Bernet and M. Ritsch-Marte, New Journal of Physics **9**, 78 (2007)
6. S. Franke-Arnold, J. Leach, M. J. Padgett, V. E. Lembessis, D. Ellinas, A. J. Wright, J. M. Girkin, P. Öhberg and A. S. Arnold, Opt. Express, **15**, 8619 (2007)
7. T. Ando, N. Matsumoto, Y. Ohtake, Y. Takiguchi, and T. Inoue, J. Opt. Soc. Am. A **27**, 2602 (2010)
8. S. Vyas, Y. Kozawa, and S. Sato, Opt. Express **21**, 8972 (2013)
9. W. Nasalski, Opt. Lett. **38**, 809-811 (2013)
10. J. A. Stratton, *Electromagnetic Theory* (McGraw-Hill, New York, 1941)
11. H. Bacry, M. Cadilhac, Phys. Rev. A 23, 2533 (1981)
12. R. W. Ziolkowski, J. Math. Phys. **26**, 861 (1985)
13. A. Sezginer, J. Appl. Phys. **57**, 678 (1985)
14. P. Hillion, J. Math. Phys. **28**, 1743 (1987)
15. A. E. Siegman, *Lasers* (University Science, Mill Valley, Calif., 1986)
16. T. Takenaka, M. Yokota, and O. Fukumitsu, J. Opt. Soc. Am. A **2**, 826 (1985)
17. W. Nasalski, Phys. Rev. E **74**, 056613 (2006)

**References with titles**

1. L. Allen, S. M. Barnett, and M. J. Padgett, Orbital Angular Momentum (Institute of Optics Publishing, 2003).
2. S. M. Barnett, Optical angular-momentum flux, J. Opt. B: Quantum Semiclass. Opt. **4**, S7-S15 (2002)
3. L. Marrucci, E. Karimi, S. Slussarenko, B. Piccirillo, E. Santamato, E. Nagali and F. Sciarrino, Spin-to-orbital conversion of the angular momentum of light and its classical and quantum applications, J. Opt. **13**, 064001 (2011).
4. M. A. A. Neil, T. Wilson and T. Juškaitis, A wavefront generator for complex pupil function synthesic and point spread function engineering, Journal of Microcopy **197**, 219-223 (2000).
5. C. Maurer, A. Jesacher, S. Fürhapter, S. Bernet and M. Ritsch-Marte, Tailoring of arbitrary optical vector beams, New Journal of Physics **9**, 78 (2007).
6. S. Franke-Arnold, J. Leach, M. J. Padgett, V. E. Lembessis, D. Ellinas, A. J. Wright, J. M. Girkin, P. Öhberg and A. S. Arnold, Optical ferris wheel for ultracold atoms, Opt. Express, **15**, 8619-8625 (2007).
7. T. Ando, N. Matsumoto, Y. Ohtake, Y. Takiguchi, and T. Inoue, Structure of optical singularities in coaxial superpositions of Laguerre- Gaussian modes, J. Opt. Soc. Am. A **27**, 2602-2612 (2010).
8. S. Vyas, Y. Kozawa, and S. Sato, Polarization singularities in superposition of vector beams, Opt. Express **21**, 8972-8986 (2013).
9. W. Nasalski, Exact elegant Laguerre-Gaussian vector wave packets, Opt. Lett. **38**, 809-811 (2013).
10. J. A. Stratton, *Electromagnetic Theory* (McGraw-Hill, New York, 1941).
11. H. Bacry, M. Cadilhac, Metaplectic group and Fourier optics, Phys. Rev. A 23, 2533-2536 (1981).
12. R. W. Ziolkowski, Exact solutions of the wave equation with complex source locations, J. Math. Phys. **26**, 861-863 (1985).
13. A. Sezginer, A general formulation of focus wave modes, J. Appl. Phys. **57**, 678 (1985).
14. P. Hillion, Spinor focus wave modes, J. Math. Phys. **28**, 1743-1748 (1987).
15. E. Siegman, *Lasers* (University Science, Mill Valley, Calif., 1986).
16. T. Takenaka, M. Yokota, and O. Fukumitsu, Propagation of light beams beyond the paraxial approximation, J. Opt. Soc. Am. A **2**, 826-829 (1985).
17. W. Nasalski, Polarization versus spatial characteristics of optical beams at a planar isotropic interface, Phys. Rev. E **74**, 056613-1-16 (2006).